\documentclass[3p,number,preprint]{elsarticle}

\usepackage[english]{babel}
\usepackage{graphicx}
\usepackage{amsmath, amsthm, amssymb, amsfonts}
\usepackage{url}

\begin{document}

\title{Adaptive logarithmic discretization for numerical renormalization
group methods}

\author[rz]{Rok \v{Z}itko}
\ead{zitko@theorie.physik.uni-goettingen.de}

\address[rz]{
Institute for Theoretical Physics, University of G\"ottingen, \\
Friedrich-Hund-Platz 1, D-37077 G\"ottingen, Germany \\
J. Stefan Institute, Jamova 39, SI-1000 Ljubljana, Slovenia
}

\date{\today}


\begin{abstract}
The problem of the logarithmic discretization of an arbitrary positive
function (such as the density of states) is studied in general terms.
Logarithmic discretization has arbitrary high resolution around some chosen
point (such as Fermi level) and it finds application, for example, in the
numerical renormalization group (NRG) approach to quantum impurity problems
(Kondo model), where the continuum of the conduction band states needs to be
reduced to a finite number of levels with good sampling near the Fermi
level. The discretization schemes under discussion are required to reproduce
the original function after averaging over different interleaved
discretization meshes, thus systematic deviations which appear in the
conventional logarithmic discretization are eliminated. An improved scheme
is proposed in which the discretization-mesh points themselves are
determined in an adaptive way; they are denser in the regions where the
function has higher values. Such schemes help in reducing the residual
numeric artefacts in NRG calculations in situations where the density of
states approaches zero over extended intervals. A reference implementation
of the solver for the differential equations which determine the full set of
discretization coefficients is also described.
\end{abstract}

\maketitle

\newcommand{\vc}[1]{\boldsymbol{#1}}
\newcommand{\ket}[1]{|#1\rangle}
\newcommand{\bra}[1]{\langle #1|}
\newcommand{\braket}[1]{\langle #1 \rangle}

\renewcommand{\Im}{\mathrm{Im}}
\renewcommand{\Re}{\mathrm{Re}}
\newcommand{\Epsilon}{\mathcal{E}}
\newcommand{\dif}{\mathrm{d}}

\bibliographystyle{unsrt} %

\section{Introduction}

Discretization of the continuum of conduction-band electron states to a
finite set of levels is a common approximation procedure in many practical
problems in computational condensed-matter physics. In band structure
calculations, for example, the discretization is performed in the reciprocal
space where a finite number of judiciously chosen (symmetry-adapted)
crystal-momentum points are chosen to sample contributions from various
regions of the full Brillouin zone \cite{monkhorst1976}. Various real-space
approaches based on finite-difference or finite-element formulations are
also possible \cite{beck2000}. Finally, in the numerical renormalization
group (NRG) methods for solving quantum impurity problems the quantity that
is discretized is the density of states, thus the discretization is
effectively performed in the energy space \cite{wilson1975, cragg1978,
krishna1980a, bulla2008}. In order to obtain good sampling of states near
the Fermi level, the discretization mesh consists of a geometric series of
points, so that very high resolution is achieved in the vicinity of the
Fermi level \cite{wilson1975}; this choice is motivated by the nature of the
quantum impurity problems (described by models such as the Kondo model or
the Anderson impurity model), where excitations from different energy scales
have a comparable effect on the physical properties. NRG allows to calculate
dynamic properties (spectral functions, dynamic susceptibilities) of
impurity models \cite{oliveira1981phaseshift, cox1985, frota1986, sakai1989,
sakai1989, yoshida1990, costi1993, costi1994, hofstetter2000, anders2005,
campo2005, peters2006, weichselbaum2007, hecht2008, resolution} and, in
particular, it may be used as an impurity solver in the dynamic mean-field
theory (DMFT) approach to the strongly-correlated electron systems
\cite{metzner1989, pruschke1995, georges1996, sakai1994, bulla1999,
pruschke2000}. Logarithmic discretization of the conductance band can also
be used in the density-matrix renormalization-group (DMRG) calculations
\cite{nishimoto2006dmrg, saberi2008, dasilva2008}, in the embedded-cluster
approximation \cite{anda2008}, and in the artificial Friedel resonance
approach \cite{bergmann2006}.

Recent work has revealed that the conventional discretization schemes used
in NRG lead to discretization artefacts, which restrain the accuracy of
calculations and limit the ultimate resolution that may be achieved
\cite{resolution}. A new approach to the discretization based on requiring
exact reproduction of the conduction-band density of states (after
discretization-mesh averaging) was shown to lead to a notable reduction of
the artefacts \cite{resolution}. In this paper the approach is further
improved by making the discretization grid itself adaptable. This work is
organized as follows: in Sec.~\ref{secproblem} the discretization problem is
formulated in very general terms starting from basic principles in order to
determine the most general form of the discretization equations. In
Sec.~\ref{secfixed}, the origin of the residual artefacts in the
fixed-discretization-mesh approach is studied, while Sec.~\ref{secadaptive}
presents a simple way of adapting the discretization grid to the density of
states. Finally, in Sec.~\ref{secode} the implementation of the
discretization equation solver is described.

\section{Problem statement}
\label{secproblem}

We consider a non-interacting Hamiltonian for the conductance-band electrons
describing a set of $N$ states indexed by some quantum number $k$:
\begin{equation}
H = \sum_{k=1}^N \epsilon_k \ket{k} \bra{k}.
\end{equation}
We assume that for all $k$, $\epsilon_k \in [-D:D]$, and in the following we
use $D$ as the unit of energy. The density of states at energy $\omega$ is
defined as
\begin{equation}
\rho(\omega) = \frac{1}{N} \sum_k \delta(\omega-\epsilon_k),
\end{equation}
and it is normalized to 1. In the continuum $N \to \infty$ limit,
$\rho(\omega)$ is an arbitrary positive function with finite support
$[-1:1]$. Our goal is to find a discrete representation of the continuum of
states that is suitable for numeric calculations in quantum impurity physics
and which reproduces the density of states $\rho(\omega)$ as accurately as
possible. The approach is clearly applicable to an arbitrary positive
function (in the single-impurity Anderson model, for example, the function
of interest is the hybridisation function $\Gamma$), but to be specific we
may think of it as a density of states.

In the following we focus on positive energies $\omega$; for negative
$\omega$ the procedure is fully equivalent. We discretize the interval
$[0:1]$ by defining the discretization-mesh points $\epsilon_j$,
$j \in \mathbb{N}$, such that
\begin{equation}
\epsilon_1 = 1, \quad
\epsilon_j > \epsilon_{j+1}, \quad
\lim_{j\to\infty} \epsilon_j = 0.
\end{equation}
In order to achieve high resolution near $\omega=0$ (i.e. in the vicinity of
the Fermi level), we furthermore require that $\epsilon_j$ behave
asymptotically as
\begin{equation}
\label{asym1}
\epsilon_j \sim \Lambda^{-j},
\end{equation}
where $\Lambda>1$ is a constant number known as the discretization
parameter. We denote each discretization interval $[\epsilon_{j+1} :
\epsilon_j]$ as $I_j$. Within each interval $I_j$ we choose a representative
energy $\Epsilon_j$. Approximating $\Epsilon_j \approx \epsilon_j$, we see
that the relative ``resolution''
\begin{equation}
\frac{\Delta E}{E} \approx \frac{E_j-E_{j+1}}{E_j+E_{j+1}}
\approx \frac{\Lambda-1}{\Lambda+1}
\end{equation}
is approximately the same on all energy scales; from this property stems the
name ``logarithmic discretization''. We furthermore notice that the
resolution is improved as $\Lambda$ is reduced toward $\Lambda=1$, which
corresponds to returning to the continuum limit. At given $\Lambda$ greater
than 1, however, the continuum spectral density is represented by a set of
delta peaks,
\begin{equation}
\label{tilde}
{\tilde \rho}(\omega) = \sum_j w_j \delta(\omega-\Epsilon_j),
\text{ for }\omega>0,
\end{equation}
where weights $w_j$ need to be chosen so that the normalization is
preserved:
\begin{equation}
\sum_{j=1}^\infty w_j = \int_0^\infty \rho(\omega)\dif \omega = W.
\end{equation}
Here we have introduced the total spectral weight for positive frequencies,
$W$. The prescription for determining the discretization-mesh points
$\epsilon_j$, representative energies $\Epsilon_j$ and weights $w_j$ (and
their negative-frequency analogues $\epsilon_j^-$, $\Epsilon_j^-$, and
$w_j^-$) is what we refer to as a ``discretization scheme''. Several such
procedures are known from the literature on the numerical renormalization
group \cite{wilson1975, frota1986, oliveira1994, chen1995, ingersent1996,
gonzalez1998, bulla1997, martinek2003b, campo2005, resolution}.

A commonly used technique to improve the accuracy of numerical calculations
is to perform computations for several different discretizations of the
continuum and average the results. In the context of the numerical
renormalization group, this is known as the ``$z$-averaging'' or the
``interleaved method'' \cite{frota1986, oliveira1994, campo2005}. In this
approach one introduces a continuous parameter $z \in [0:1]$ which
characterizes different discretizations $\{ \epsilon_j^z, \Epsilon_j^z,
w_j^z \}$. Parameter $z$ is sometimes called the ``twist parameter'' since
it can be related to twisting the boundary conditions of the wave-functions
of the conduction-band electrons \cite{campo2005}. The $z$-averaging
procedure is meaningful when the boundary conditions 
\begin{equation}
\label{boundary}
\forall z\ \, \epsilon_1^z = 1, \quad
\Epsilon_1^0 = 1,
\end{equation}
and continuity constraints 
\begin{equation}
\label{cont} 
\forall j\ \, \epsilon_j^1 = \epsilon_{j+1}^0, \quad
\forall j\ \, \Epsilon_j^1 = \Epsilon_{j+1}^0,
\end{equation}
are satisfied.
We also require that $\epsilon_j^z$ and $\Epsilon_j^z$ be monotonously
decreasing functions of $z$, but they need not be continuous. Furthermore,
for each value of $z$, the weights need to be normalized:
\begin{equation}
\label{norm0}
\sum_{j=1}^\infty w_j^z = W.
\end{equation}
The spectral density is then determined as the average (integral) over
$z$ \cite{campo2005}:
\begin{equation}
A(\omega) = \int_0^1 {\tilde \rho}(z,\omega)\,\dif z,
\end{equation}
where ${\tilde \rho}(z,\omega)$ is given by the generalization of
Eq.~\eqref{tilde}:
\begin{equation}
{\tilde \rho}(z,\omega) = \sum_j w^z_j \delta(\omega-\Epsilon^z_j).
\end{equation}
We obtain \cite{campo2005}
\begin{equation}
\label{spec}
A(\omega) = \sum_j \int_0^1 w_j^z \delta(\omega-\Epsilon_j^z)\,\dif z
= \frac{w_j^z}{-\partial E_j^z/\partial z},
\end{equation}
where the parameters $j$ and $z$ in the last part of the expression are
determined implicitly through the relation $\omega=\Epsilon_j^z$; this
choice of $j$ and $z$ is unique due to the requirement that $\Epsilon_j^z$
is strictly decreasing as a function of $z$. In practical calculations, only
a small number $N_z$ of values of $z$ is used (often as small as 4 or 2, or
even a single one) and the integral is approximated using the rectangle
method; the resulting function then needs to be broadened appropriately to
obtain a continuous representation. See, however, Ref.~\cite{resolution} for
calculations with a large $N_z$ and very narrow broadening kernel, which
show that overbroadening errors of NRG can largely be eliminated.

To simplify the notation and to provide more insight into the mathematical
structure of the problem, we introduce continuous indexing of
the discretization-grid points as \cite{resolution}
\begin{equation}
x=j+z.
\end{equation}
The ``grid parameter'' $x$ runs from $1$ to $+\infty$ and the coefficients
$\epsilon_j^z$, $\Epsilon_j^z$ and $w_j^z$ become continuous functions
$\epsilon(x)$, $\Epsilon(x)$ and $w(x)$, while the continuity constraints,
Eq.~\eqref{cont}, are automatically satisfied. Both $\epsilon(x)$ and
$\Epsilon(x)$ are required to be monotonously decreasing and, furthermore,
the following boundary conditions need to be satisfied:
\begin{eqnarray}
\epsilon(x) = 1 \text{ for } x \leq 2, \quad
\lim_{x\to\infty} \epsilon(x) = 0, \\
\Epsilon(1) = 1, \quad
\lim_{x\to\infty} \Epsilon(x) = 0.
\end{eqnarray}
These equations embody the requirement that as the grid parameter $x$ sweeps
the interval $[1:+\infty)$, both the grid energy $\epsilon(x)$ and the
representative energy $\Epsilon(x)$ describe the totality of the unit
interval $[0:1]$. For convenience, we define the inverse function $R$ to
$\Epsilon$ as $x=R(\omega)$, i.e. $R \circ \Epsilon = 1$.

Using newly introduced notation, Eq.~\eqref{spec} becomes
\begin{equation}
\label{Aomega}
A(\omega)=\frac{w(x)}{-\dif\Epsilon(x)/\dif x}, \text{ with } x=R(\omega).
\end{equation}
The normalization condition can be expressed as
\begin{equation}
\label{norm1}
\int_0^1 A(\omega)\,\dif\omega = \int_\infty^1 A(\omega)
\frac{\dif \Epsilon(x)}{\dif x} \,\dif x = \int_1^\infty w(x)\,\dif x = W,
\end{equation}
but, in addition, the more general normalization equation
\begin{equation}
\label{norm2}
\sum_{j=1}^{\infty} w(j+z)=W,
\end{equation}
which follows from Eq.~\eqref{norm0}, must be satisfied for each $z \in
[0:1]$; in fact, Eq.~\eqref{norm1} follows trivially from Eq.~\eqref{norm2}.
We notice that the function $\epsilon(x)$ appears neither in
Eq.~\eqref{Aomega} nor in Eq.~\eqref{norm2}; grid coefficients $\epsilon(x)$
are thus only auxiliary quantities which define the discretization grid
without actually explicitly appearing in the final result for the density of
states.

With all these preparations we can now finally state the problem as follows:
we seek to determine functions $\Epsilon(x)$ and $w(x)$ such that
\begin{equation}
\label{problem}
\frac{w[R(\omega)]}{-\dif \Epsilon[R(\omega)] / \dif
x}=\rho(\omega),
\end{equation}
and satisfying normalisation, Eq.~\eqref{norm2}.

Since, $\epsilon(x)=1$ for $x \in [1:2]$ and $\epsilon(x)\to 0$ as
$x\to\infty$, it is easy to see that $w(x)$ defined as
\begin{equation}
w(x)=\int_{\epsilon(x+1)}^{\epsilon(x)} \rho(\omega)\,\dif\omega
\end{equation}
solves the normalization Eq.~\eqref{norm2}. In fact, this is the only
general solution of Eq.~\eqref{norm2}; we note, however, that $\epsilon(x)$
can in principle still be an arbitrary monotonously decreasing function on
the interval $[2:\infty)$.  Thus $w(x)$ {\sl must} be defined as the
integral of the density of states in the discretization integral
$I(x)=[\epsilon(x+1) : \epsilon(x)]$, while we still have the full liberty
of choosing the discretization mesh in any convenient way.

To make contact with the original motivation for introducing the logarithmic
discretization of a continuum, we now focus on discretization meshes
$\epsilon(x)$ with asymptotic behavior $\epsilon(x) \sim \Lambda^{2-x}$,
where we have shifted the exponent by 2 for convenience. This asymptotic
form is equivalent to requiring
\begin{equation}
\label{eqe}
\frac{\dif\epsilon(x)}{\dif x} = -\Lambda^{2-x}\,\ln \Lambda\, C(x, \epsilon),
\end{equation}
where $C(x,\epsilon)$ is an arbitrary strictly positive function with
a non-zero limit 
\begin{equation}
\lambda = \lim_{\substack{x \to \infty \\ \epsilon \to 0}} C(x,\epsilon).
\end{equation}
The requirement $\lambda \neq 0$ is necessary to obtain the desired
asymptotic behavior.

The problem is thus reduced to finding an appropriate function $C(x,
\epsilon)$. Once $C(x, \epsilon)$ is chosen, the full solution of the
problem is obtained by solving the initial value problem
\begin{equation}
\label{ode1}
\begin{split}
\Epsilon(1) &= 1, \\
\frac{\dif\Epsilon(x)}{\dif x} &= -\frac{\int_{\epsilon(x+1)}^{\epsilon(x)}
\rho(\omega)\dif\omega}{\rho[\Epsilon(x)]}.
\end{split}
\end{equation}

In NRG, the discrete levels resulting from the discretization for given
twist parameter $z$ are used to write the discretized form of the
conduction-band Hamiltonian
\begin{equation}
H=\sum_{i=1}^{\infty} \Epsilon_i^z \ket{i}\bra{i}.
\end{equation}
One then forms the combination of levels
\begin{equation}
\ket{f_0} = \sum_i \sqrt{w_i} \ket{i},
\end{equation}
and transforms the Hamiltonian $H$ into a new basis (the first state of
which is $\ket{f_0}$), so that the Hamiltonian takes the form of a
tight-binding chain \cite{wilson1975, krishna1980a, bulla2005, bulla2008}:
\begin{equation}
\label{Hband}
H=\sum_{n=0}^\infty \xi_n \ket{f_n}\bra{f_n} + \sum_{n=0}^{\infty}
t_n \left( \ket{f_n}\bra{f_{n+1}} + \ket{f_{n+1}}\bra{f_n} \right).
\end{equation}
This form is known as the ``hopping Hamiltonian'' or the ``Wilson chain''.
The coefficients $t_n$ decrease asymptotically as $t_n \propto
\Lambda^{-n/2}$ and only a finite number of chain sites is retained in
practical NRG calculations. The impurity hybridizes with the conduction band
through level $\ket{f_0}$ only; this implies that the spectral function
$A_f(\omega)$ of the level $f_0$ represents the density of states of the
conduction band as seen by the impurity. The goal is thus to make
$A_f(\omega)$ a good representation of the density of states $\rho(\omega)$.

We will discuss possible choices of $C(x,\epsilon)$ in
Sec.~\ref{secadaptive} and describe a numerical approach for solving the
initial value problem in Sec.~\ref{secode}, but we first pause to describe
the previously known discretization schemes based on fixed discretization
mesh to point out the origin of their deficiencies.

\section{Fixed discretization mesh}
\label{secfixed}

The conventional logarithmic discretization used in NRG calculations is
based on the fixed discretization mesh obtained by setting $C(x,\epsilon)
\equiv 1$ in Eq.~\eqref{eqe}, with solution
\begin{equation}
\epsilon(x) = 
\begin{cases}
1  & \text{ for } x \in [1:2],\\
\Lambda^{2-x} & \text{ for } x \geq 2.
\end{cases}
\end{equation}
The different discretization schemes then differ in the recipe for
calculating $\Epsilon(x)$. The first schemes for an arbitrary density of
states \cite{chen1995, ingersent1996, bulla1997} used a physically
motivated, but otherwise ad-hoc expression
\begin{equation}
\Epsilon(x) = \frac{\int_{\epsilon(x+1)}^{\epsilon(x)} \rho(\omega)\,
\omega\,\dif\omega}
{\int_{\epsilon(x+1)}^{\epsilon(x)} \rho(\omega)\,\dif\omega},
\end{equation}
i.e. an average of energy weighted by $\rho(\omega)$. An improved approach
was later introduced \cite{campo2005}, where
\begin{equation}
\Epsilon(x) = \frac{\int_{\epsilon(x+1)}^{\epsilon(x)} \rho(\omega)\,
\dif\omega}
{\int_{\epsilon(x+1)}^{\epsilon(x)} \rho(\omega)/\omega\,\dif\omega}.
\end{equation}
This scheme has better convergence properties as $\Lambda \to 1$ and it
doesn't systematically underestimate the density of states at low energies.
Neither of these two approaches, however, in general satisfies
Eq.~\eqref{problem}, which leads to artefacts which become apparent in
high-resolution NRG calculations, in particular at high energies near the
band edges \cite{resolution}. More recently, a scheme based on solving
Eq.~\eqref{ode1} for a fixed discretization mesh was shown to be very
successful in removing the most severe of these artefacts \cite{resolution}.
In most cases, this approach works well, it is robust and the
conductance-band density of states can be reproduced accurately in numerical
calculations. We find, however, that in practical NRG calculations some
residual artefacts still appear, in particular in situations where the
density of states has large and rapid variations. These artefacts cannot be
fully removed by increasing the number of $z$-values used in the averaging.
The artefacts appear especially near energies $\Epsilon_j^1$, $j \in
\mathbb{Z}$. Their origin can be traced back to the systematic errors in NRG
calculations, which shift the spectral peaks from exact energies
$\Epsilon_j^z$ to $\Epsilon_j^z + \delta \Epsilon_j^z$. For given $j$, the
error $\delta \Epsilon_j^z$ is a smooth function of parameter $z$, thus
$\delta \Epsilon_j^{n/N_z}-\delta \Epsilon_j^{(n+1)/N_z}$ is a small
quantity for all $n=1,{\ldots},N_z-1$. Even a narrow broadening kernel will
then give a smooth final result in these energy ranges and the resolution
can in principle be systematically improved by decreasing the broadening
width while simultaneously increasing the total number of $z$ values, $N_z$.
To the contrary, the difference in the values of $\delta
\Epsilon_j^{N_z/N_z}$ and $\delta \Epsilon_{j+1}^{1/N_z}$ is large, since
the two calculations are based on two very different discretization meshes.
This is still true even for very large $N_z$. Near $\Epsilon_j^1$ points,
thus, there exist a minimum broadening width below which these systematic
errors become unmasked and manifest themselves as sharp artefacts. This
effectively limits the highest spectral resolution that one may expect to
achieve in NRG calculations.

It was found that problems of this type become especially pronounced for
densities of states which are very low (or zero) over considerable energy
intervals. Such situation commonly occurs when NRG is used as the impurity
solver in the dynamical mean-field theory \cite{sakai1994, bulla1999,
pruschke2000}. It should be noted that a representative state will be chosen
in each interval even if the interval contains very small (or even zero)
spectral weight. While such states appear with little weights $w_i$, their
presence is nevertheless found to be detrimental to the overall accuracy of
the calculation. The issue is, actually, somewhat subtle: the weights $w_i$
only affect the combination of states that forms $\ket{f_0}$. The terms
$\Epsilon_i \ket{i}\bra{i}$ in regions of small density of states still
appear in $H$ of Eq.~\eqref{Hband} and they affect the coefficients $\xi_n$,
$t_n$ of the Wilson chain in a way which is detrimental for the accuracy of
the calculation. This problem is especially severe for very small $z$, where
some of the coefficients $\xi_n$ may become very large (when expressed in
units of $\Lambda^{-n/2}$).

We now study the severity of this problem on the example of a density of
states with a smooth transition from lower density at high energies to
higher density at lower energies. As a simple model we choose a density of
states described by
\begin{equation}
\rho(\omega) \propto 1+w \tanh\left[-a(|\omega|-\omega_0)\right],
\end{equation}
where $w$ ($0 \leq 0 \leq 1$) determines the height of the variation,
$\omega_0$ the energy where the change occurs and $a$ its rapidity.

\begin{figure}
\centering
\includegraphics[width=10cm]{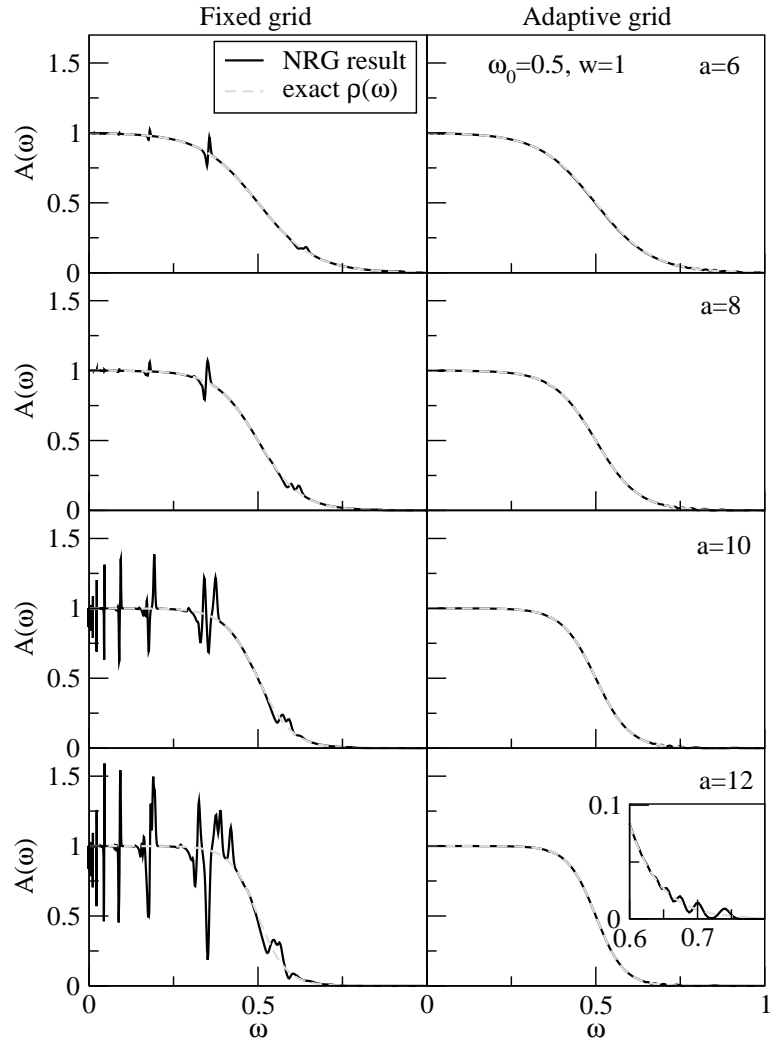}
\caption{Spectral function of the first site on the Wilson chain
for $\tanh$ density of states of the conduction band: comparison
between results using fixed and adaptive discretization grids
for a range of parameter $a$, which controls the steepness of
the transition and consequently the value of $\rho(\omega)$
in the low-density range.}
\label{figa}
\end{figure}

\begin{figure}
\centering
\includegraphics[width=10cm]{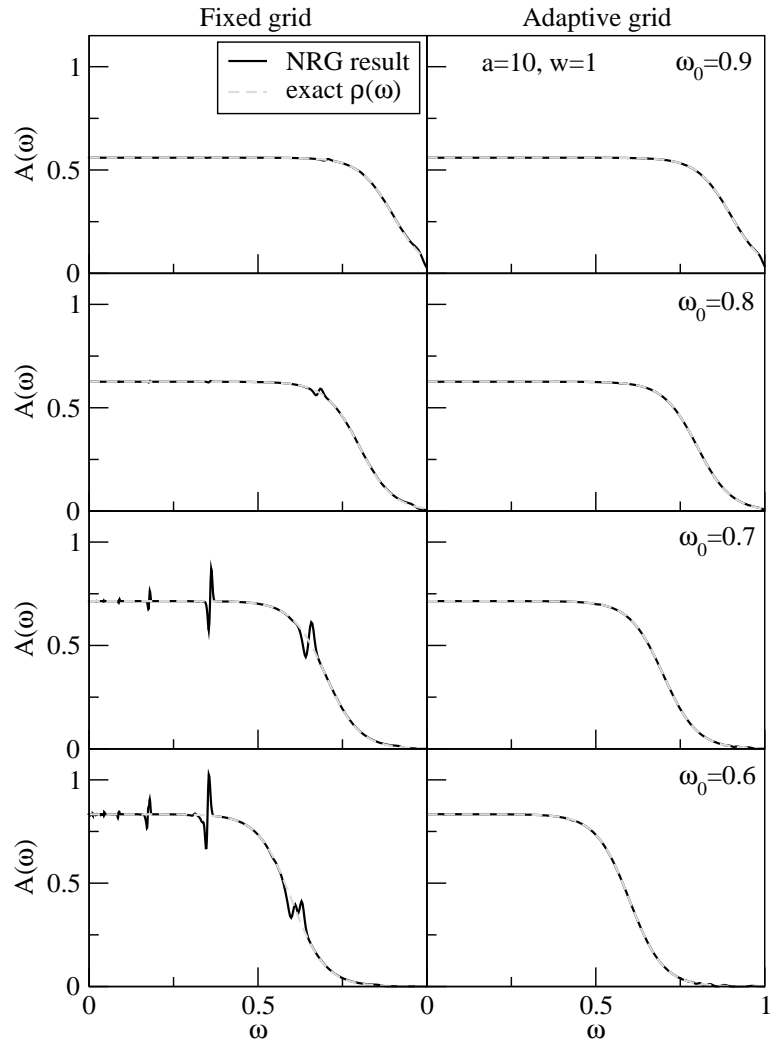}
\caption{Spectral function of the first site on the Wilson chain for $\tanh$
density of states of the conduction band: comparison between results using
fixed and adaptive discretization grids for a range of parameter $\omega_0$,
which controls the energy of the transition from high-density to low-density
range.} \label{figb}
\end{figure}

In Figs.~\ref{figa} and \ref{figb} we compare $A_f(\omega)$, the spectral
function on the first site of the Wilson chain (level $f_0$), computed using
the NRG, with the model density of states $\rho(\omega)$. Calculations were
performed with $N_z=64$ $z$ values, discretization parameter $\Lambda=2$,
truncation cutoff set at $10\omega_N$, broadening parameter $\eta=0.01$ and
patching parameter $p=2.1$ (for details on the method see
Ref.~\cite{resolution}). We focus now on left panels which contain results
obtained using fixed grid discretization; right panels with adaptive-grid
calculation results will be discussed in Sec.~\ref{secadaptive}. In
Fig.~\ref{figa} the transition is made progressively sharper. At first only
the artefacts at energies $\Epsilon_j^1$ are amplified, however for large
enough $a$ additional sharp features appear. In Fig.~\ref{figb}, we shift
the transition point to lower energies, thereby again increasing the range
of very low density of states. While for $\omega_0=0.9$ the artefacts are
very mild, they increase significantly as soon as $\rho(\omega)$ becomes
very low for large $\omega$. It may be noted that mild artefacts may easily
be removed by making the broadening kernel wider, but one then needs to
renounce on obtaining high-resolution spectral functions. When artefacts are
severe, they might affect even calculations with very wide broadening
kernel.

\section{Adaptive discretization mesh}
\label{secadaptive}

From the considerations detailed in the previous section it follows that a
more appropriate choice of the discretization grid should be such that the
grid were less dense in the regions of low density of states. A convenient
way to implement this requirement is by demanding that the function
$C(x,\epsilon)$ in Eq.~\eqref{eqe} be of the form
\begin{equation}
C(x,\epsilon) = \frac{f(x)}{\rho(\epsilon)}.
\end{equation}
It can be seen that in the limit $\rho(\epsilon) \to 0$, the corresponding
energy regions will not even appear in the discretization grid, which is a
very desirable property.

We consider a specialization to $f(x) \equiv A$, where $A$ is a constant. It
is easy to see that $A$ must, in fact, be equal to the total weight
$W=\int_0^1 \rho(\omega)\,\dif\omega$. We also observe that for a
featureless flat band [$\rho(\omega)=1/2$], this approach reduces to the
conventional fixed discretization grid.

We furthermore note that the ``characteristic energy scale at the $N$th NRG
iteration'', $\omega_N$, depends on the discretization grid (i.e. on
function $\epsilon(x)$), thus it becomes dependent on $\rho(\omega)$. This
has important consequences on the choice of parameters in NRG calculations;
in particular, the truncation cutoff and the patching parameter $p$ need to
be appropriately redefined. A simple choice is to rescale $\omega_N$ by the
ratio
\begin{equation}
R=\lim_{x\to\infty} \frac{\epsilon(x)}{\epsilon_0(x)},
\end{equation}
where $\epsilon_0(x)$ is the discretization grid function for fixed grid.
This choice appears suitable for the $\tanh$ test functions, but it is not
expected to be generally applicable. For strongly varying density of states,
the patching procedure itself might become problematic and it might be
better to use the complete-Fock-space approach \cite{anders2005, peters2006,
weichselbaum2007}, although that technique is not without problems either
\cite{resolution}. The issue of extracting spectral information from partial
results at different iterations clearly merits further attention.

Comparing the right panels in Figs.~\ref{figa} and \ref{figb} with the left
panels, we see that the adaptive-grid approach leads to a considerable
improvement; the artefacts essentially disappear. We notice that some
oscillations appear in the transition region between low and high spectral
density, see the inset in Fig.~\ref{figa}. It is important to notice that if
the discretization mesh is adaptive, the width of the broadening kernel in
the calculation of the continuous spectral function should ideally also take
the mesh density into account; energy regions where mesh is less dense
should be broadened more, since there will be less representative energy
points. At the same time, one should be careful to avoid excessive
broadening of spectral functions on the impurity levels, since low density
in the conductance band translates into sharper features in the impurity
spectral function. We note that in the extreme case of a gap in the
conductance band, there might even appear delta peaks in the impurity
spectral function. One should thus base the choice of the broadening width
on physical considerations and on the expected spectral features.

\section{Initial value problem}
\label{secode}

Both $\epsilon(x)$ and $\Epsilon(x)$ must have asymptotic behavior
$\Lambda^{2-x}$; this is required to obtain a logarithmic discretization
around the point $x=0$. For numerical solution of the equations,
it is therefore convenient to use the following Ansatz:
\begin{equation}
\epsilon(x) = g(x) \Lambda^{2-x}, \quad
\Epsilon(x) = f(x) \Lambda^{2-x}.
\end{equation}
The unknown functions $g(x)$ and $f(x)$ are then ${\mathcal O}(1)$
for all $x$. The differential equations to be solved are
\begin{equation}
\begin{split}
\frac{\dif g(x)}{\dif x} &= \ln \Lambda \left\{ g(x) - 
C\left[x,g(x) \Lambda^{2-x}\right] \right\}, \\
\frac{\dif f(x)}{\dif x} &= \ln \Lambda f(x) -
\frac{\int_{g(x+1)\Lambda^{1-x}}^{g(x)\Lambda^{2-x}} \rho(\omega)\,\dif\omega}
{\Lambda^{2-x} \rho[ f(x) \Lambda^{2-x} ]},
\end{split}
\end{equation}
with initial conditions $g(2)=1$ and $f(1)=1/\Lambda$. One first solves for
$g(x)$, which is then used in the equation for $f(x)$. Both equations are
stiff, thus some care is needed in their numeric solution to prevent that
$g(x)$ or $f(x)$ diverge.

When NRG is used as an impurity solver in the dynamical mean-field theory,
the input to the calculation is the effective impurity hybridisation
function which contains information on the density of states of the
self-consistently defined medium. The hybridisation function usually takes
the form of a tabulated function. A differential equations solver was
implemented which solves for $g(x)$ and $f(x)$ on a mesh of values of $x$,
given an arbitrary input density of states (or hybridisation function). The
software package is available from the author's home page
(\url{http://nrgljubljana.ijs.si/adapt}). For better portability the solver
is written in pure ISO C++ without making use of any external libraries.

In the solver, $\rho(\omega)$ is calculated for an arbitrary point $\omega$
from the tabulated input function by linear interpolation. Integrations are
performed using trapezoidal method; where the integration boundary falls
inside a tabulation interval the contribution is, however, calculated by
explicitly integrating the linear interpolation function. For numerical
stability, it is very important that interpolation and integration be of
consistent order of approximation. The differential equations are solved
using fourth-order Runge-Kutta solver with adaptive reduction of the
integration step size near the boundaries of the tabulation intervals. This
is especially important when solving for $f(x)$ in situations where
$\rho(\omega)$ has sharp features (steps, kinks, sharp peaks). In the
reference implementation of the solver, we use
$C(x,\epsilon)=A/\rho(\epsilon)$. Constant $A$ is first estimated by the
numerical integral $W=\int_0^1 \rho(\omega)\dif\omega$, but it is then
refined by repeatedly solving the differential equation for $g(x)$ until
$g(x)$ no longer tends to diverge; to determine a suitable value $A$ in this
shooting-method approach we use the secant method to solve the equation
\begin{equation}
g(x_\mathrm{max}) = \frac{A}{\rho(0)}.
\end{equation}
When solving the differential equation for $f(x)$, it is observed that the
integration steps need to be made small enough in regions of $x$ which
correspond to varying $\rho(\omega)$, otherwise $f(x)$ will diverge. For
large $x$ in the region where $\rho[f(x) \Lambda^{2-x}]$ is essentially
constant, longer steps may be used.

The output from the program are tabulated functions $f(x)$ and $g(x)$ which
can then be read as input to the NRG code.

\section{Conclusion}

A possible improvement would consist of choosing such function
$C(x,\epsilon)$ to make the discretization grid denser in the regions of
large variation of $\rho(\epsilon)$, in other words to make it also depend
on the derivative $\dif\rho/\dif\epsilon$. Alternatively, one could
systematically study the influence of various environmental modes (i.e. sets
of nearby conduction-band states) on system dynamics and choose
$C(x,\epsilon)$ so that sampling is denser in energy regions which
contribute more \cite{zwolak2008}. Furthermore, the possibility of making
the function $C$ depend on the grid parameter $x$ could be used to deform
the mesh so that the first weight $w_1^z$ would not be small for small $z$;
$w_1^z$ tends to 0 with $z \to 0$ even if the density of states near the
band edges is large, which leads to similar problems as discussed in
Sec.~\ref{secfixed}, although with lesser severity.

The author acknowledges discussions with Thomas Pruschke, computer support
by GWDG and support by the German Science Foundation through SFB 602.

\bibliography{paper}

\end{document}